\documentclass{aa}
\usepackage{graphicx}
\usepackage{natbib}
\bibpunct{(}{)}{;}{a}{}{,} 
\usepackage{txfonts}
\begin{document}
   \title{Morphology and evolution of umbral dots and their substructures}
   
   \author{M. Sobotka
          \inst{1}
          \and
          K. G. Puschmann
          \inst{2}
          }

   \offprints{M. Sobotka}

   \institute{Astronomical Institute, Academy of Sciences of the Czech Republic
              (v.v.i.), Fri\v cova 298, 25165 Ond\v rejov, Czech Republic\\
              \email{msobotka@asu.cas.cz}
         \and
             Instituto de Astrof\'\i sica de Canarias, V\'ia L\'actea S/N,
             38200 La Laguna, Tenerife, Spain\\
             \email{kgp@iac.es}
             }

   \date{Received April 21, 2009; accepted June 15, 2009.}
   
   \abstract
   {Substructures -- dark lanes and tails -- of umbral dots (UDs) were
predicted by numerical simulations of magnetoconvection and have been
detected later in some observations.}
   {To provide constraints for realistic theoretical models of
sunspot umbrae, we describe the observed properties and evolutionary
characteristics of UDs (including their substructure) and of other umbral
structures.}
   {We analyse a 6~h~23~min time series of broadband images of a large
umbra in the active region NOAA 10634, acquired with the 1-m Swedish
Solar Telescope, in the wavelength band around 602~nm. A 43~min part
of this series was reconstructed with the MFBD method,
reaching a spatial resolution of 0\farcs14. With the help of image
segmentation, feature tracking, and local correlation tracking, we measure
brightness, size, lifetime, and horizontal velocities of various umbral
structures.}
   {Large structures in the umbra -- strong and faint light bridges
(LBs) and an extended penumbral filament -- evolve on time scales of hours.
Most (90~\%) of UDs and bright point-like features in faint LBs split
and merge, and their median lifetimes are 3.5 or 5.7~min, depending on whether
the split or merge event is considered as the end of their life. Both UDs and
features in faint LBs that do not split or merge are clearly smaller
(0\farcs15) than the average size (0\farcs17) of all features.
Horizontal motions of umbral bright small-scale features are directed
either into the umbra or along faint LBs with mean horizontal velocities
of 0.34\,km~s$^{-1}$. Features faster than 0.4\,km~s$^{-1}$
appear mostly at the periphery of the umbra. The motion of peripheral UDs
(PUDs) seems to be the continuation of the motion of penumbral grains (PGs).
The intensity of dark lanes, measured in four bright central UDs (CUDs),
is by a factor 0.8 lower than the peak intensity of CUDs. The width of
dark lanes is probably less than the resolution limit 0\farcs14.
The characteristic time of substructure changes of UDs is $\sim$ 4 min.
We observe narrow (0\farcs14) bright and dark filaments connected 
with PUDs. The bright filaments are 0.06 $I_{\rm ph}$ brighter than 
the dark ones. Usually one dark and two bright filaments form a 0\farcs4 wide 
tail attached to one PUD, resembling a short dark-cored penumbral filament.}
   {Our results indicate the similarity between PUDs and PGs located at
the tips  of bright penumbral filaments. The features seen in numerical 
MHD simulations are consistent with our observations of dark lanes in CUDs 
and tails attached to PUDs.}

   \titlerunning{Morphology and evolution of umbral dots}

   \maketitle
%

\section{Introduction}\label{sec:intro}

The study of fine structure in sunspot umbrae is crucial to understanding
the interactions of plasma with strong magnetic fields and the physical
structure of sunspots. Currently, thanks to new ground-based and space
instruments, image reconstruction methods, and MHD simulations,
our knowledge of fine-structure elements is showing considerable
progress. This concerns umbral dots (UDs), the small isolated bright features
observed in the umbra, and light bridges (LBs), the elongated bright features
crossing the umbra. At the umbra-penumbra boundary, penumbral grains (PGs)
located at the tips of bright penumbral filaments sometimes protrude deep
into the umbra. Usually, UDs are divided into two groups, peripheral umbral
dots (PUDs) and central umbral dots (CUDs), according to their position
inside the umbra \citep{grdoerth:86}. For definitions and basic
characteristics of the sunspot fine structures, we refer to the monograph
by \citet{thomweiss:08}.

Two basic approaches have been proposed to explain the nature of UDs.
According to \citet{choudhuri:86}, UDs are columns of hot field-free gas
intruding into a cluster of thin magnetic flux tubes that form the magnetic
structure of a sunspot \citep{parker:79}. Another approach, based on the
magnetoconvection in a monolithic thick flux tube, has been proposed by
\citet{weissetal:90}. In their model, UDs are hot convective plumes that
overshoot to the sunspot's photosphere. This approach was extensively
elaborated by means of 3D numerical simulations of magnetoconvection.
Recently, \citet{schussvog:06} have shown that the energy transport in
the vertical magnetic field of the umbra is dominated by nonstationary
narrow plumes of rising hot plasma with adjacent downflows. The magnetic
field is strongly reduced in the upper layers of the plumes. At the
visible surface ($\tau = 1$), the plumes are seen as bright features with
a typical size of 200--300 km and lifetime of the order of 30 minutes,
resembling UDs. Most of the simulated UDs have an elongated shape with a
central dark lane, some larger UDs show a threefold dark lane. The dark lanes
are a consequence of enhanced density and opacity at the top of the plume,
so that the continuum formation layer is shifted upwards to cooler layers
of the atmosphere. \citet{rempel:09} extended the simulation, introducing
a 2D (slab) magnetic field fanning out with height into a 3D convective
media. They obtained a sunspot-like structure surrounded by photospheric
granulation. In the umbra of the ``sunspot'', UDs similar to those simulated
by \citet{schussvog:06} were present. At the periphery of the umbra in
a more inclined magnetic field (with respect to the vertical), UDs
appeared with two attached bright narrow tails pointing toward the
penumbra. Even farther from the centre of the umbra structures developed
resembling PGs followed by dark-cored penumbral filaments
\citep[see Fig.~1 of ][]{rempel:09}.

The most recent extensive imaging observations of UDs have been published by
\citet{riethm:08b}. They observed thousands of UDs in a 106~min long time
series of images acquired in the TiO band head (705.7 nm) with the
1-m Swedish Solar Telescope (SST). Applying a feature tracking algorithm,
they measured sizes, brightnesses, lifetimes, trajectories, and horizontal
velocities. The majority of UDs was spatially resolved with typical
diameter of 0\farcs 31. A difference in the evolution of brightness is
found between CUDs and PUDs, pointing to a different origin of these
types of UDs. The behaviour of CUDs shows better agreement with the
simulations of \citet{schussvog:06} than that of PUDs. A new subset of
PUDs, mobile UDs with travel distances over 1\arcsec, was introduced.
Mobile UDs are born close to the umbra-penumbra boundary, they are brighter,
live longer, and move faster (680 m~s$^{-1}$) than the rest of UDs
(410 m~s$^{-1}$). Properties of UDs were also recently studied by
\citet{kitai:07} in three series of broad-band images acquired by Hinode/SOT.

Recent spectroscopic and spectropolarimetric observations have provided
new information about line-of-sight (LOS) velocities and magnetic fields.
Analysing Dopplergrams obtained at the Dunn Solar Telescope,
\citet{bharti:07a} found upflows of 400 m~s$^{-1}$ surrounded by downflows
of 300 m~s$^{-1}$ in large (0\farcs 5) UDs. The discovery of downflows
strongly supports the magnetoconvective origin of UDs.
Full Stokes spectra of 21 CUDs and 30 PUDs, observed by Hinode/SOT, were
inverted by \citet{riethm:08a}. At the continuum level ($\tau = 1$),
they find a magnetic field weakening of 480 G in CUDs and 510 G in PUDs.
No significant LOS velocities have been detected in CUDs, but PUDs are connected
with upflows of 800 m~s$^{-1}$. The physical differences between CUDs and
PUDs have been confirmed by \citet{sobjur:09}, using a 3~h long time series
of full-Stokes spectral scans obtained with Hinode/SOT. In the range of
optical depths $0.3 < \tau <0.6$, CUDs do not show any excess of LOS
velocity and magnetic field inclination with respect to the surrounding
umbra, while upflows of 400 m~s$^{-1}$ and a more horizontal magnetic
field are observed in PUDs. It seems that PUDs are more like PGs
than CUDs. Both PUDs and PGs appear in regions with a weaker and more
horizontal magnetic field and their formation height reaches the low
photosphere, while CUDs appear in regions with stronger and more vertical
magnetic fields, and they are formed too deep to detect upflows and changes
in magnetic field inclination.

Dark lanes in UDs, predicted in the simulations by \citet{schussvog:06},
have been observed by \citet{bharti:07b} in large ($>$1\arcsec) bright UD-like
features seen in Hinode/SOT \mbox{G-band} images with diffraction-limited spatial
resolution of 0\farcs 22. After inverting a nearly contemporaneous Hinode/SOT
spectropolarimetric scan, \citet{bharti:09} conclude that the dark lanes
correspond to a temperature deficit located above $\tau = 0.1$, which,
together with observed upflows and magnetic field reduction in UDs, supports
the results of theoretical simulations. In \mbox{G-band} images acquired with
the Dunn Solar Telescope, \citet{rimmele:08} observed dark dots or lanes
within bright UDs. Their size is typically close to the diffraction
limit of 0\farcs 12. Close to the umbral border, Rimmele finds elongated
dark structures associated with bright PUDs moving into the umbra. These
dark structures have the appearance of a short dark tail pointing toward
the penumbra, and they resemble dark-cored filaments originating at PGs.

In this paper we describe the fine structure of a large and dark umbra,
observed for more than 6 hours with the spatial resolution 0\farcs 14
in a large number of broad-band images. First we describe the observed
long-term characteristics and evolution of the umbra,
LBs, and long penumbral filaments extending into the umbra.
Then we centre on some statistical properties of umbral features.
Some preliminary results  were published previously in two conference
papers \citep[][ hereafter Papers I and II]{sobpusch:07,sobetal:08}.
Finally we focus on the substructures of UDs -- dark lanes in CUDs and
fine bright and dark filaments composing tails associated with PUDs.

\section{Observations and data processing}\label{sec:obs}

The large leading sunspot in NOAA 10634 was observed on 18 June 2004
from 07:43 to 15:30 UT with the 1-m SST, La Palma \citep{schar:2003},
equipped with adaptive optics. The spot, located near the disk centre
(N13, E12), was growing to the maximum phase of evolution.
The broad-band images were acquired simultaneously in three wavelength
bands: blue ($450.8 \pm 0.5$) nm, red ($602.0 \pm 1.3$) nm, and \mbox{G-band}
($430.89 \pm 0.6$) nm. Exposure times were 11--14 ms and the pixel size
was 0\farcs 0405 $\times$ 0\farcs 0405.
The images were acquired in the frame-selection mode during most of the
observing period. The selection interval was 20 s and the average time
difference between two frames (including the time necessary to store
the images) is 20.1~s. From 08:40 to 09:30 UT, all frames
(cadence 4--5 s$^{-1}$) were stored to be used later for image
reconstruction by MFBD (the special treatment of these data is
described in this section later on). A post-facto frame selection
was applied to these frames to obtain a homogeneous time series during
the full observing period with a temporal resolution of 20.1~s.

After dark- and flatfield corrections, the intensities in each image
were normalised with respect to the average intensity $I_{\rm ph}$ of
a quiet granulation area near the sunspot to compensate for changes
of transparency and/or exposure time.
Photospheric and penumbral stray light, which contaminates the umbra
due to the scattering in the terrestrial atmosphere and in the instrument,
was eliminated using the method by \citet{marpil:1992}. Parameters of
the scattering were determined for each wavelength from the shapes of
photometric profiles across the solar limb. Images of the limb were
recorded immediately after the observation. The level of stray light,
expected to originate mostly in the instrument, was 8.5 \% in blue,
6.5 \% in red, and 8.7 \% in G-band. A deconvolution of the instrumental
profile of the diffraction-limited 1-m telescope and a simultaneous
noise filtering was applied to the frames of the series, using
Wiener filters with noise suppression starting at 0\farcs 11 (blue),
0\farcs 14 (red), and 0\farcs 13 (G-band). Regarding the correction
of wavefront aberrations done by the adaptive optics, these values
characterise the spatial resolution in the best frames. Then,
the image rotation was compensated and the frames were aligned and
de-stretched. Finally, a subsonic filter with a cutoff at 4 km s$^{-1}$
was applied to remove a residual jitter due to the seeing. The
beginnings and ends of the series were removed because of insufficient
image quality and apodisation effect. The resulting three time series
in blue, red, and G-band wavelength ranges span from 08:05 to 14:28 UT,
i.e., 6 h 23 min. The G-band series is not suitable for the study
of umbral structures, because the darkest parts of the umbra are
underexposed. The red continuum series with the best signal-to-noise
ratio ($S/N = 200$) was selected for further analysis.

As mentioned above, a part of the time series (from 08:40 to 09:30 UT)
has been treated with special care. During this 50 min time period
the seeing was not extremely good but very stable, offering a good
opportunity for an efficient image restoration. We stored all the images
(a total of 12000 frames in each blue and red band) acquired at a frame
rate of 4--5 images~s$^{-1}$. We applied the post-facto multi-frame blind
deconvolution (MFBD) image restoration algorithm developed and implemented
by \citet{loefdahl:02} on the recorded dark- and flat-field corrected
images. This reconstruction method uses multiple frames
to retrieve the residual aberrations affecting the images
after the partial correction performed by the adaptive optics and thus
further improves the image quality. In the present case we combined sets
of 50 sequential images to obtain single restored ones. Thus, the resulting
sequences of the red and blue bands last 50 min and consist of 236
reconstructed images with a temporal cadence of 12.6~s each. This sampling
interval has been chosen such that the amount of images grouped for the
restoration is as large as possible without losing important physical
information about the temporal evolution of our science target. Moreover,
the reconstruction is performed in small portions of the images where
one can consider a common characteristic aberration (isoplanatic patches).
Overlapping patches of 128$\times$128 pix (5\farcs184$\times$5\farcs184)
have been extracted such that at the end of their restoration the whole
field of view (FOV) can be recovered by mosaicking them without any
discontinuity. Proceeding in this way, the homogeneity and image quality
along the time series and in the whole FOV can be improved up to close
to the diffraction limit of the telescope. The resulting reconstructed
time series have been treated with the standard procedures, i.e.,
image de-rotation, image alignment, and subsonic filtering (again with
a cutoff of 4 km s$^{-1}$). Some frames of these series were lost due to
apodisation, so that the resulting 43~min long blue- and red-band series 
contain 204 frames each and span from 08:43 to 09:26 UT.

The horizontal motions of umbral features were studied using the methods
of local correlation tracking \citep[LCT,][]{novsim:1988} and feature
tracking. LCT provides a time-averaged horizontal velocity field of all
structures in the FOV. The FWHM of the Gaussian tracking window
was set to 0\farcs 32 and the temporal integration was made over 1~h and
2~h. There was no substantial difference between the velocity fields
obtained with these two integration times. The feature tracking provides
time records of positions, intensities, and sizes for each individual
feature. We used the algorithm developed by \citet{sbs:1997} that has
recently been improved by \citet{vafa:08}. This algorithm makes it possible
to distinguish between features that split into two or more parts, merge
with another feature, or preserve their integrity during their life.

\section{Results}\label{sec:res}

\subsection{Long-term evolution of umbral structures}\label{subsec:evol}

\begin{figure}
  \centering
  \includegraphics[width=0.48\textwidth]{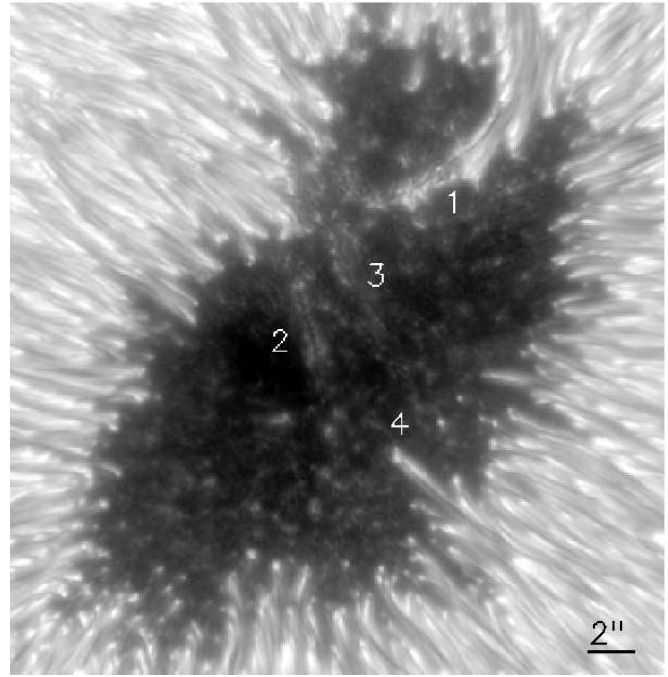}
  \caption[]{\label{fig:1}
Red-band image of the umbra taken at 12:21 UT.
1 -- strong light bridge; 2,3 -- faint light bridges;
4 -- long penumbral filament.
}
\end{figure}

The evolution of the umbra was inspected visually in the movie composed
of the red-band images taken from 08:05 to 14:28 UT. One of the images,
taken at 12:21 UT, is shown in Fig.~\ref{fig:1}.
The umbra was very stable during our observation. Its area
$A = 345$~arcsec$^2$, corresponding to the effective diameter
$d_{\rm eff} = \sqrt{4\,A/\pi} = 21$\arcsec,  was constant within
this period. The area was defined by the intensity threshold
0.5~$I_{\rm ph}$. The observed umbra is very dark. After stray-light
correction, the minimum  intensity in the umbra is 0.09~$I_{\rm ph}$
(0.05~$I_{\rm ph}$ in the blue band), remaining constant with time.
The whole umbra shows a slowly evolving grainy and/or filamentary
low-contrast ``background'' pattern with rms fluctuations of only
0.01--0.02 $I_{\rm ph}$ (Paper I).
Bright inward-moving UDs travel in this background pattern. Even a 
dark nucleus, the less intense part of the umbra, where the strongest 
magnetic field is expected, shows a very faint grainy and filamentary
structure.  Identical patterns were detected in the red and blue bands,
proving that we observe real structures, so the umbral background does
not have the diffuse character expected in the past.

\begin{figure}
  \centering
  \includegraphics[width=0.48\textwidth]{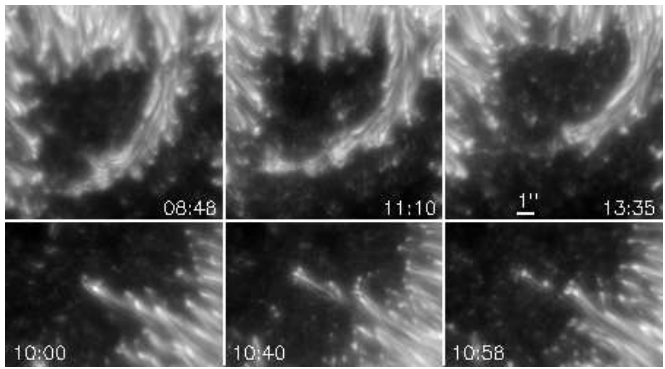}
  \caption[]{\label{fig:snap}
Snapshots of evolution of the strong light bridge marked as ``1''
in Fig.~\ref{fig:1} (top row) and of the long penumbral filament
``4'' (bottom row).
}
\end{figure}

A strong LB, marked as ``1'' in Fig.~\ref{fig:1}, divided the umbra into
two umbral cores. One part of the bridge (top right in Fig.~\ref{fig:1})
was composed of extended penumbral filaments with dark cores, while the
other part had a granular structure with a central dark lane. The dark
lane was not connected to any of the dark cores in the penumbral filaments.
After 3~h~5~min of the observation (at 11:10 UT), the granular part split
into several fragments that weakened gradually, until the minimum
brightness was reached at 13:35 UT (see snapshots of evolution in
Fig.~\ref{fig:snap}, top row). Then the granular part started to recover.
Bright grains moved along the filamentary part of the LB toward the
granular one with typical velocities of 0.8--1 km~s$^{-1}$ (LCT) all
the time.

The long penumbral filament ``4'', extending deep into the umbra,
passed several evolutionary phases. It had been present since the beginning
of the observation. Starting from 08:22 UT, its length increased with
an average speed of 0.6 km~s$^{-1}$ during 40 min. After that growing
phase, it remained stable for 50 min and then it started to grow
again. The second growing phase took approximately 30 min. In the next
20 min the filament separated from the penumbra and then, after a
30 min long decay phase, it disappeared at 11:10 UT. The bright head
of the filament converted into three PUDs that moved toward LB~2.
Three snapshots displayed in the bottom row of Fig.~\ref{fig:snap}
show this filament in the second growing phase, during its separation
from the penumbra, and in the decay phase.
At the same time (11:10 UT), a new long penumbral filament developed
at the distance of 1\farcs 3 from the original one. It showed the same
phases of evolution at very similar time scales -- growth (50 min,
speed 0.6 km~s$^{-1}$), stability (50 min), second growth (30 min),
separation (20 min), and decay (30 min). It disappeared at 14:11 UT
and its total lifetime was 3~h. During the existence of both filaments,
bright PGs moved along them with speeds up to 1.2 km~s$^{-1}$ (LCT) and
many PUDs, moving toward LB~2, detached from the heads of the filaments.

\begin{figure}
  \centering
  \includegraphics[width=0.48\textwidth]{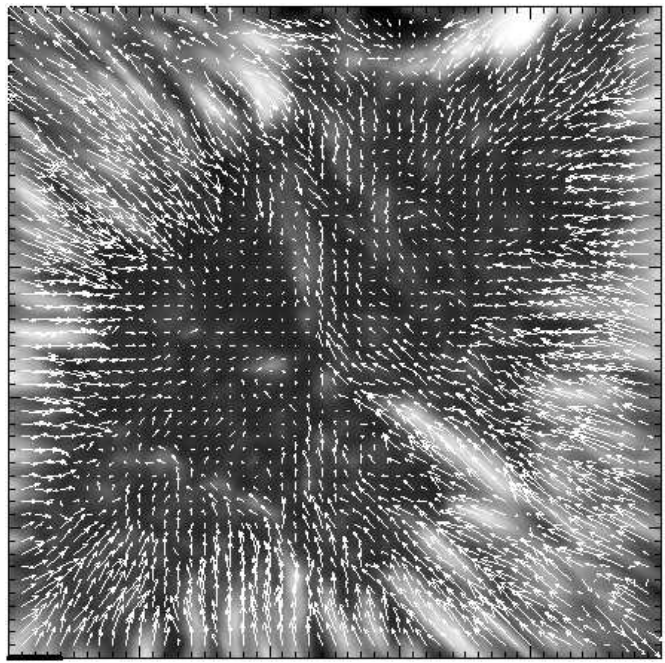}
  \caption[]{\label{fig:LCT}
LCT map of horizontal velocities, calculated with the tracking window
of 0\farcs 32 and averaged over a period of 117 minutes, the best part
of the unreconstructed series in the red band, taken in the
frame-selection mode from 12:15 to 14:12 UT. The length of the black
horizontal bar at the bottom left corner represents 1~km~s$^{-1}$.
}
\end{figure}

Faint granular LBs ``2'' and ``3'' (Fig.~\ref{fig:1}) started to form
at 10:05 and 09:15 UT, respectively. They developed into
well-distinguishable structures with central dark lanes around 11:45 UT.
During the formation period, the motion (0.2--0.3 km~s$^{-1}$, LCT)
of bright features along LB~3 and along the upper part of LB~2 was
oriented into the umbra. In the lower part of LB~2, an opposite motion 
toward the penumbra (0.3 km~s$^{-1}$) was seen. When the formation was
finished, the opposite motion of 0.4 km~s$^{-1}$ prevailed in LB~2
(see Paper I).
Such opposite motions along two adjacent LBs were also observed by
\citet{rimmele:08}. This LB started to decay at 13:55 UT,
while LB~3 remained stable till the end of the observation. The opposite
motion of bright features in LB~2 might be related to a horizontal
velocity field connected with two long penumbral filaments that
developed subsequently at the position ``4'' in Fig.~\ref{fig:1}.
An example of an LCT map of horizontal velocities in the umbra
is shown in Fig.~\ref{fig:LCT}. This map is calculated for the period
from 12:15 to 14:12 UT.

Bright inward-moving PUDs are often followed by faint bright filaments,
resembling penumbral filaments attached to PGs, and/or by dark ``wakes'',
resembling dark cores of penumbral filaments (Paper I). Similar structures
were reported independently by \citet{rimmele:08}. The LCT flow
map (see Fig.~\ref{fig:LCT}) shows that the horizontal motion of PUDs
(0.5 km~s$^{-1}$) seems to be the continuation of the inward-directed 
motion field of PGs in the inner penumbra.

\subsection{Statistical properties of umbral features}\label{subsec:stat}

In the following we focus on some detailed statistical results obtained
for the best part of the unreconstructed series in the red band, taken
in the frame-selection mode from 12:15 to 14:12 UT. Movies covering
90 minutes of the time series from 12:28 to 13:58 UT can be found at
{\small {\tt http://www.asu.cas.cz/$\sim$sdsa/ gallery-astropictures.html}}.

The feature-tracking technique was applied to this 117 min long
unreconstructed time series. The bright features were
separated from the background using the low-noise curvature determination
algorithm \citep{vafa:08}. Only features (UDs and dot-like features in
faint LBs) brighter by 0.04~$I_{\rm ph}$ than the surrounding umbra,
larger than 0\farcs 14 (resolution limit), and living longer than 2~min
were taken into account. The resulting sample contains 3099 UDs and 1357
dot-like features in LBs.

The majority (90 \%) of observed features split
or merge with other features (see Paper II), thus it is necessary to take
into account these effects. One possibility is to finish the time record
of a feature when it splits or merges. In this case, the mean (median)
lifetimes of UDs are 4.5~min (3.7~min) and the mean (median) lifetimes of
the features in faint LBs are 4.4~min (3.4~min). On the other hand, split
and merge events caused by seeing effects and random coincidences may
spuriously shorten the lifetime of many UDs. The used feature tracking
algorithm makes it possible to connect the records of features before
and after the split/merge event according to the following rules:
(a) if a feature splits, its successor is the brightest fragment;
(b) if two or more features merge, the resulting feature is connected
to its longest-lived predecessor. In this case, the mean (median)
lifetimes of UDs are 9.1~min (5.7~min) and the mean (median) lifetimes
of the features in faint LBs are 9.8~min (5.7~min). Independently of
the applied method, the histograms of lifetimes show a monotonous
decrease from shortest to longest lifetimes as reported previously
by \citet{sbs:1997} and \citet{riethm:08b}.

\begin{figure}
  \centering
  \includegraphics[width=0.48\textwidth]{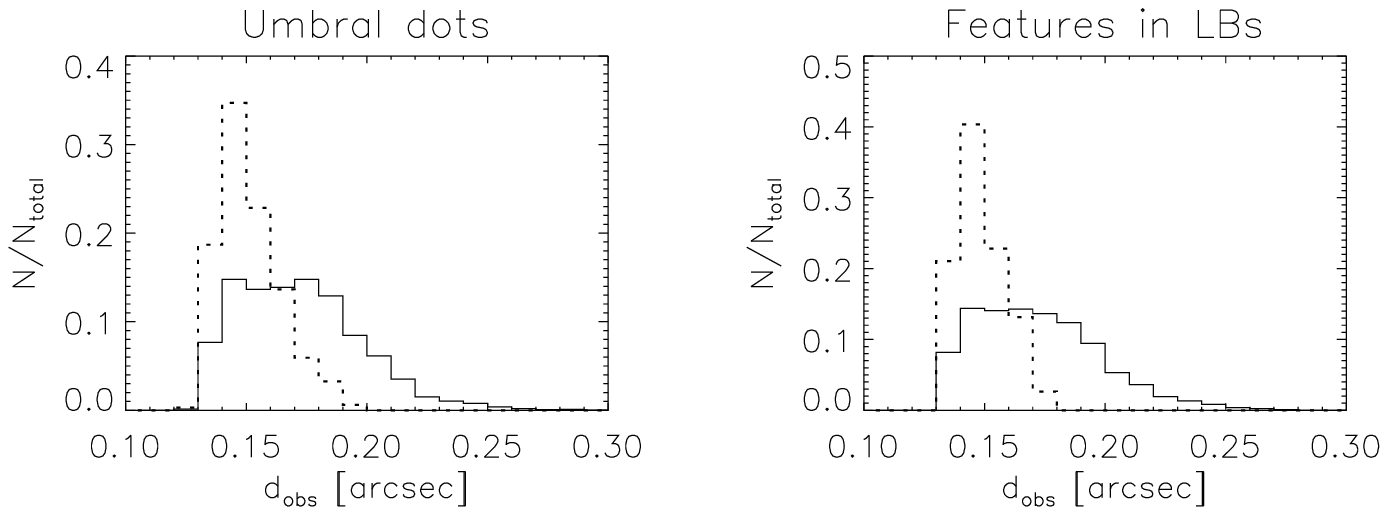}
  \includegraphics[width=0.48\textwidth]{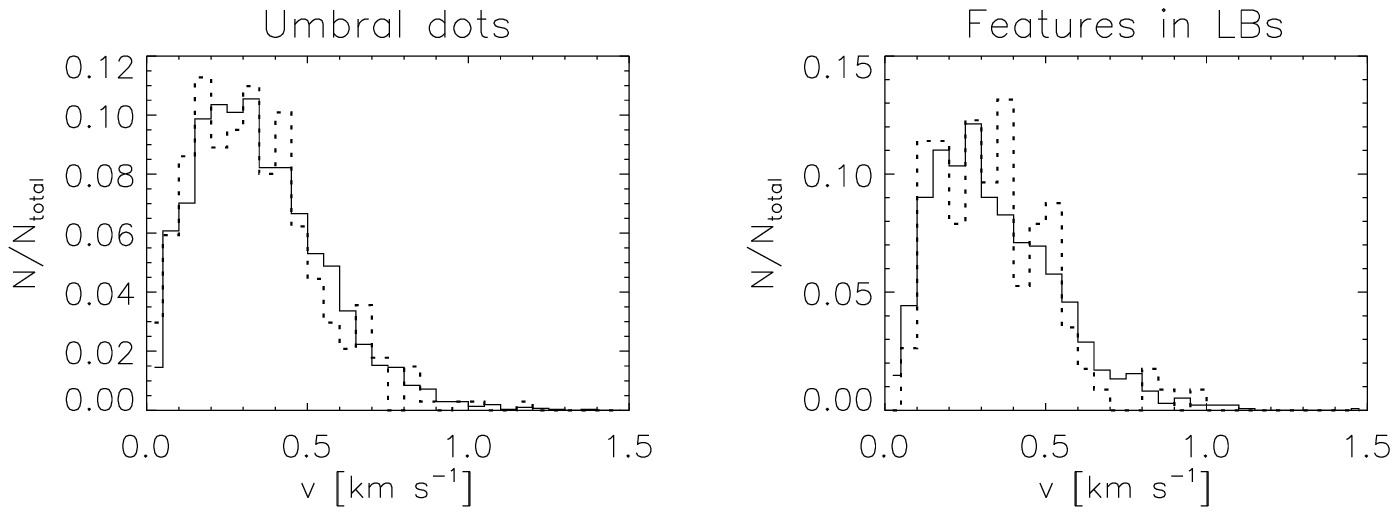}
  \caption[]{\label{fig:kinema4}
Histograms of time-averaged effective diameters (top) and
time-averaged horizontal velocity magnitudes (bottom), derived
from feature tracking. Dotted line -- non-split/merge features,
solid line -- all features.
}
\end{figure}

Both UDs and features in faint LBs~2 and 3 that do not split
or merge are smaller than the average size of all features
(Paper II). Histograms of effective diameters calculated from
time-averaged areas of all 3099 UDs and 1357 bright features in LBs
and of 337 UDs and 114 bright features in LBs that did not
split or merge are plotted in Fig.~\ref{fig:kinema4},
top row. It can be seen that UDs and bright features in LBs do not differ
in size. However, the non-split/merge features have narrower range of sizes
and smaller mean size (0\farcs 15) than all features in the sample
(with the mean size of 0\farcs 17). Larger diameters of splitting or
merging features suggest that these features may be composed to
a large extent of small unresolved objects.

The feature tracking was also used to measure the horizontal
velocities of small-scale umbral features. Motions of UDs are mostly
directed into the umbra, while the features in faint LBs~2 and 3
move predominantly along the bridges.
Histograms of time-averaged velocity magnitudes are plotted in
Fig.~\ref{fig:kinema4}, bottom row. We see
that the velocity magnitudes are similar for UDs and features in LBs
and do not depend on split or merge of the features. All types of
features show a mean velocity around 0.34 km~s$^{-1}$.
A map of horizontal velocity vectors (not shown here) indicates that
features faster than 0.4 km~s$^{-1}$ appear mostly at the periphery
of the umbra and in faint LBs, where the magnetic field is expected
to be weaker and more horizontal (see Paper II).

\begin{figure}
  \centering
  \includegraphics[width=0.48\textwidth]{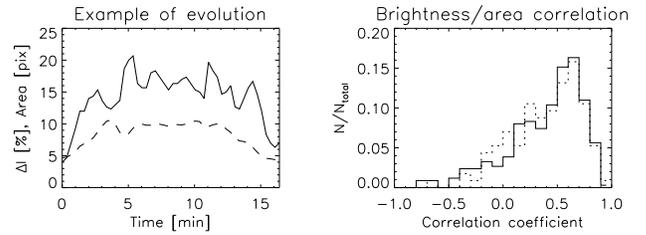}
  \caption[]{\label{fig:kinema5}
Left -- example of area (solid) and brightness (dashed)
evolution of a central umbral dot. Right -- histograms of
the correlation between brightness and area during the evolution
of individual non-split/merge features. Solid line -- umbral dots,
dotted line -- bright features in faint light bridges.
}
\end{figure}

The time-averaged values of brightness, size, and horizontal velocity
of small-scale umbral features are completely uncorrelated.
However, during the evolution of an individual feature, there is a
correlation between the brightness excess $\Delta I$ with respect to
the surrounding umbra and the area of the feature (Paper II). An example
of the evolution of a long-lived CUD is shown in the left panel of
Fig.~\ref{fig:kinema5}. The correlation between $\Delta I$ and area was 0.68
in this case. Coefficients of correlation between $\Delta I$ and area
evolution were calculated for the non-split/merge features (337 UDs and
114 bright features in faint LBs~2 and 3) and the histograms are plotted in
the right panel of Fig.~\ref{fig:kinema5}. The temporal brightness and size
variations are positively correlated in most cases -- 50 \% of features
have correlation coefficients higher than 0.48 and the typical value
of a correlation is 0.60. The features first grow and brighten and then
get smaller and fainter, in agreement with theoretical predictions
\citep[e.g.][]{schussvog:06}.
For the other features that split or merge, the $\Delta I$-to-area
correlation is much weaker but still detectable. The histograms (not
shown here) are asymmetric with prevailing positive values and have flat
maxima around 0.40. For 50 \% of split/merge features, the correlation
coefficients are higher than 0.22. The reason for the weak correlation
is that the split/merge events introduce substantial uncorrelated
changes in brightnesses and sizes of the features.

\subsection{Dark lanes in central umbral dots}\label{subsec:dls}

\begin{figure}
  \centering
  \includegraphics[width=0.48\textwidth]{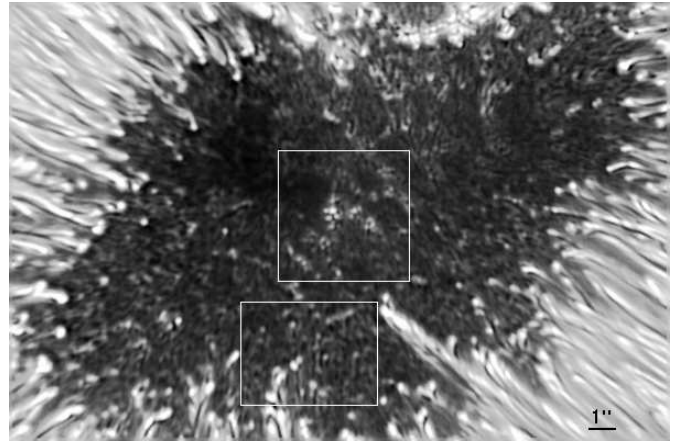}
  \caption[]{\label{fig:2}
MFBD reconstructed red-band image of the umbra taken at 08:59 UT.
This image is enhanced by the unsharp masking technique for
a better visualisation.
The upper rectangle contains a group of central umbral dots
with dark lanes, the lower one a field of peripheral umbral dots
with tails.
}
\end{figure}

A group of bright (0.3--0.5 $I_{\rm ph}$), large (0\farcs4), and
long-lived CUDs was present in the central part of the umbra from
08:30 to 10:05 UT (see Fig.~\ref{fig:2}, upper box). This group moved
slowly in the direction of LB~2. Dark lanes, predicted by theoretical
simulations \citep{schussvog:06}, are observed in several CUDs of this
group. Obviously, because of the limited spatial resolution, we are able
to observe dark lanes only in bright and large UDs. According to
simulations, such UDs should show multifold dark lanes, separating
them into several parts. In real observations, it is difficult
to distinguish such UDs from conglomerates of unrelated small UDs.
However, the integrity of the observed CUDs during the time period
of about one hour indicates that we indeed see CUDs with dark lanes.

The 43 min series of MFBD reconstructed red-band frames from 08:43 to
09:26 UT with a temporal resolution 12.6 s is used to study the
substructures in the four brightest CUDs in detail. Equivalent
substructures are also seen in the reconstructed blue-band frames,
but with lower signal-to-noise ratio. Three- or fourfold dark lanes
separate CUDs into 3--4 parts. The substructures vary with time (the
typical time scale is about 3~min) and the dark lanes disappear
and reappear during the evolution of CUDs.
These changes may be caused by residual seeing variations.

\begin{figure}
  \centering
  \includegraphics[width=0.48\textwidth]{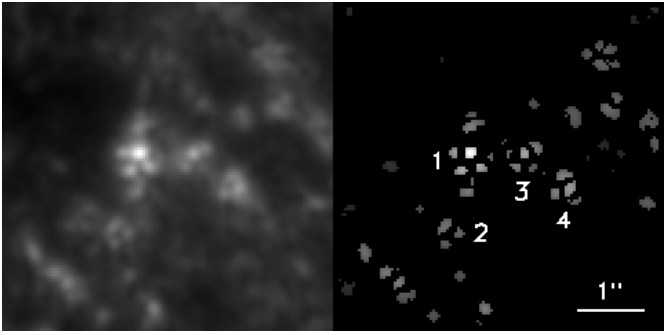}
  \caption[]{\label{fig:3}
Central umbral dots with dark lanes. The field corresponds to the
upper box in Fig.~\ref{fig:2}. Left -- original image, right --
corresponding segmented image. Umbral dots under study are marked
by numbers.
}
\end{figure}

The measurement of brightness, size, and temporal changes of four CUDs,
their parts, and dark lanes was made in a series of segmented images
of the field delimited by the upper box in Fig.~\ref{fig:2}. The
segmentation was made in two steps. First, CUDs were separated
from the background using the multilevel segmentation
\citep[][T. Roudier 2003, private communication]{bovelet:01}. Then
the dark lanes and parts of CUDs were identified by means of
the low-noise curvature determination algorithm \citep{vafa:08}.
An example of the original and segmented image is displayed in
Fig.~\ref{fig:3}.

The basic observed parameters of CUDs 1--4 are summarised in
Table~\ref{tab:1}. The table contains the values of CUD peak intensity
$I_{\rm ud}$, intensity $I_{\rm b}$ of the surrounding umbral
background, intensity $I_{\rm dl}$ of the dark lanes, and diameter $d$.
All the intensities are in units of $I_{\rm ph}$. The values of intensities
and diameters are the averages of measurements in four frames for each CUD
showing the best-resolved substructure. The lifetime $t$ of CUDs
is measured in the 6~h~23~min long time series and $t_{\rm dl}$ is the
total time for which the dark lanes were observed during the 43~min
long series of reconstructed frames. The peak-to-background ratio
$I_{\rm ud}/I_{\rm b}$ for the brightest CUD 1 is 3.9, for CUDs 2--4
is approximately 2.5. The average ratio $I_{\rm dl}/I_{\rm ud}$ is 0.8.
Since the dark lanes are very narrow, this value has to be considered
as an upper limit.

\begin{table}
\caption{Observed parameters of central umbral dots with dark lanes
(see text for the explanation of symbols).}
\label{tab:1}
\centering
\begin{tabular}{ccccccc}
\hline\hline
\noalign{\smallskip}
CUD & $I_{\rm ud}$ & $I_{\rm b}$ & $I_{\rm dl}$ & $d$ & $t$ [min] & $t_{\rm dl}$ [min] \\
\hline
\noalign{\smallskip}
 1      &  0.54  &  0.14  &  0.43  & 0\farcs 41 &  80  &  35  \\
 2      &  0.36  &  0.15  &  0.28  & 0\farcs 40 &  75  &  10  \\
 3      &  0.43  &  0.18  &  0.32  & 0\farcs 42 &  70  &  14  \\
 4      &  0.43  &  0.16  &  0.37  & 0\farcs 44 &  50  &  14  \\
\hline
\end{tabular}
\end{table}

The feature-tracking algorithm was applied to the 43 min
series of segmented images to measure the lifetimes and sizes of the
individual parts of CUDs, separated by the dark lanes. The mean and
median lifetimes are 9 min and 4 min, respectively.
The median value is comparable to the characteristic time of
temporal variations of the substructures (3~min) estimated by the
visual inspection of the movie. Residual seeing fluctuations
may be the cause of these changes.
Both the mean and median sizes of the parts of CUDs are 0\farcs 17.
The apparent width of the dark lanes measured in the segmented
images is 0\farcs 08--0\farcs 12 (2--3 pixels), i.e., smaller than
our resolution limit 0\farcs 14. Nevertheless, because the dark lanes
are easily detected in the restored images (see Fig.~\ref{fig:3}, left
panel), we may expect that they are not much narrower than 0\farcs 14.

\subsection{Tails of peripheral umbral dots}\label{subsec:tails}

\begin{figure}
  \centering
  \includegraphics[width=0.48\textwidth]{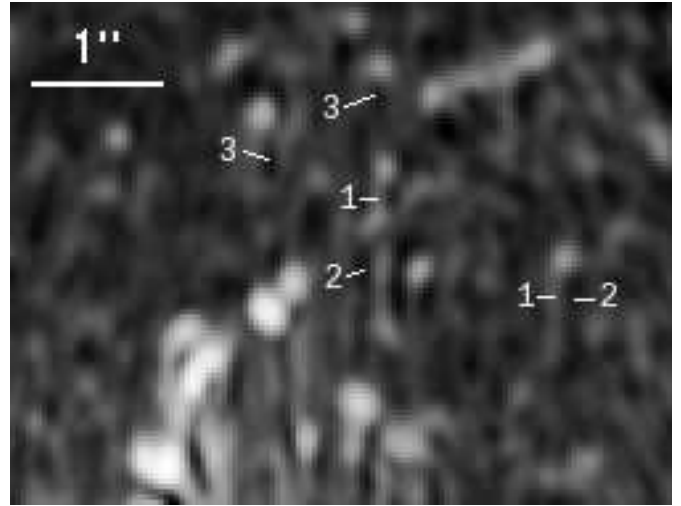}
  \caption[]{\label{fig:4}
Narrow bright and dark filaments (tails)
connected to peripheral umbral dots. Numbers label examples of:
1 -- bright filaments, 2 -- dark filaments, 3 -- tails composed
of two bright and one dark filament. The image is enhanced by
the unsharp masking technique for a better visualisation.
The field corresponds to the lower box in Fig.~\ref{fig:2}.
}
\end{figure}

In Paper I we mentioned that narrow bright or dark tails are often
attached to PUDs. In the 43~min series of MFBD reconstructed red- and
blue-band images, a well-resolved pattern of radially oriented narrow
bright and dark filaments is observed in the peripheral parts of the
umbra (see Fig.~\ref{fig:2}). These filaments are not connected to the
penumbra. The rms contrast of the pattern is only 0.02--0.03 $I_{\rm ph}$
in the red-band images. Most of the filaments are attached to PUDs,
giving the appearance of comet-like tails pointing in the direction
toward the penumbra \citep[cf.][]{rimmele:08}. The presence of the
attached filaments is independent of the brightness of PUDs. An example
is shown in Fig.~\ref{fig:4}, a magnification of the lower box in
Fig.~\ref{fig:2}. The image is enhanced by unsharp masking for a better
visualisation of the small-scale structure. Examples of bright
and dark filaments are labelled by numbers 1 and 2, respectively.

The basic observed parameters of the bright and dark filaments were
measured in the best images of the reconstructed red-band series.
The typical width of the filaments is at or below the resolution
limit 0\farcs 14. Their length ranges from 0\farcs 4 to 1\arcsec.
The intensity of the bright filaments is approximately 0.06 $I_{\rm ph}$
higher than the intensity of the dark ones. The filaments live
5--20~min, what is comparable with the lifetime of UDs.

In many cases, two bright filaments with a dark one between them
are attached to one PUD, forming a composite tail approximately
0\farcs 4 wide, similar to a short dark-cored penumbral filament.
Such dark-cored tails are seen in Fig.~\ref{fig:4} and also in
the right part of Fig.~\ref{fig:2}
at the periphery of the umbra and near the strong light bridge.
Sometimes we observe that one bright filament is missing, probably
because it is too faint and narrow to be detected, sometimes the
dark-cored tails overlap. For this reason we expect that the
dark-cored tails of PUDs are in fact more numerous than we observe.
Such composite tails also appear as a result of the theoretical
simulations of magnetoconvection in an inclined magnetic field.
They are well visible in Fig.~1 of \citet{rempel:09} at the
periphery of the simulated umbra.

\section{Discussion and conclusions}\label{sec:disc}

We analysed the 6~h~23~min time series of broadband images
($\lambda = 602.0 \pm 1.3$ nm) of a large ($d_{\rm eff} =  21$\arcsec)
and dark ($I_{\rm min} = 0.09\ I_{\rm ph}$) umbra. A 43~min part
of this series was reconstructed by means of the MFBD method
\citep{loefdahl:02} and used to study the substructures of
UDs -- dark lanes in CUDs and narrow tails attached to PUDs.

Two faint LBs with dark central channels (LB~2 and LB~3, see
Fig.~\ref{fig:1}) were formed in the umbra. The formation phase took
about 2~h, followed by the phase of stability. Small (0\farcs 17)
bright dot-like features moved along these LBs. The dark central
channels in faint LBs may have a similar physical origin like the
dark lanes in UDs, i.e., the enhanced density at the top of ascending
hot gas \citep{rimmele:08,giordano:08}.
Two bright penumbral filaments extending deep into the umbra developed
subsequently at nearby positions (``4'' in Fig.~\ref{fig:1}). Each one
showed two phases of growth (50 and 30 min), then a separation from the
penumbra (20 min), and a final decay into several PUDs (30 min). The
total lifetime of one penumbral filament was 3~h.

Lifetimes, sizes, and horizontal velocities of UDs and
dot-like features in faint LBs were measured using a feature tracking
technique applied to the 117 min long best part of the unreconstructed
time series. The majority (90 \%) of all observed features split
or merge with other features. The median lifetime of both types of
features is approximately 3.5~min if the life of a feature is expected
to finish when it splits or merges. If the split/merge events are not
considered to be the end of life (according to the rules described in
Section~\ref{subsec:stat}), the median lifetime is 5.7 min.
Both UDs and features in faint LBs that do not split or merge are
clearly smaller (0\farcs 15) than the average size (0\farcs 17)
of all features. This means that features showing the split or merge
events may be composed to a large extent of small unresolved objects.
The obtained sizes are much smaller compared to 0\farcs 31 reported
by \citet{riethm:08b}. Horizontal motions of umbral bright
small-scale features are directed either into the umbra or along
faint LBs. Time-averaged velocity magnitudes (0.34 km~s$^{-1}$)
are similar for UDs and features in faint LBs and do not depend
on the split or merging of the features. Features faster than
0.4 km~s$^{-1}$ mostly appear at the periphery of the umbra
and in faint LBs, where the magnetic field is expected to be weaker
and more horizontal. Although the time-averaged values of
brightness, size, and horizontal velocity of features are uncorrelated,
during the evolution of individual non-split/merge features the temporal
brightness variations are positively correlated with the size variations
in most cases.

For the first time, we tried to measure the detailed photometric
parameters of CUDs with dark lanes in the series of reconstructed
images. For this purpose we selected a group of four bright, large,
and long-lived CUDs in the centre of the umbra.
They were brighter by a factor of 2.5 (3.9 in one case) than
the surrounding background. The three- or fourfold dark lanes,
observed with the resolution of 0\farcs 14, correspond to narrow
-- below the resolution limit -- 20~\% deep depressions in the
intensity profiles of CUDs. This result has to be taken with care,
because of the strong contamination by stray light from the bright
parts of CUDs. The substructures vary strongly with time.
The median lifetime of the bright parts of CUDs is 4 min, and the dark
lanes disappear and reappear. It is a question of whether these changes,
not seen in the simulations by \citet{schussvog:06}, might be caused by
the residual seeing variations in the reconstructed images. The average
size and lifetime of the four selected CUDs with dark lanes (0\farcs 42
and 70 min) are comparable to the corresponding parameters of UDs
appearing in the simulations (0\farcs 3--0\farcs 4 and 30 min).
However, such large and long-lived CUDs are not frequent in umbrae.
The majority of UDs are smaller (0\farcs 2--0\farcs 3) and live much
shorter time (3--9 min) -- see also \citet{riethm:08b}. The future
theoretical simulations should reproduce also such small and
short-lived UDs.

We confirm the bright and dark filamentary tails connected
to PUDs and directed toward the penumbra, reported in Paper I and by
\citet{rimmele:08}. In fact, such substructures are attached to nearly
all PUDs, independent of their brightness. We newly observe that
many tails of PUDs are composed of two narrow bright and one dark
parallel filaments, forming a structure 0\farcs 4 wide that resembles
a short dark-cored penumbral filament. The length of this
structure is 0\farcs 4--1\arcsec. The lifetime of such dark-cored
tails is comparable to the lifetime of UDs. It can be expected that
the origin of central dark filaments in the tails is analogous to the
origin of the dark lanes in UDs and dark cores in penumbral filaments.
\citet{sobjur:09} found that, concerning the physical characteristics,
PUDs are similar to PGs located at the tips of bright penumbral filaments.
The discovery of the dark-cored tails attached to PUDs strongly supports
this similarity. Such structures also appear as a result of the
radiative MHD simulation of a sunspot by \citet{rempel:09}. These
authors claim that the underlying processes for the formation of
penumbral filaments are the same as those for umbral dots. From this
point of view, PUDs with dark-cored tails result from magnetoconvective
energy transport in the form of hot rising plumes inside the inclined
magnetic field at the periphery of the umbra. They represent a transition
from CUDs to PGs and bright penumbral filaments.

We conclude that the numerical MHD simulations of magnetoconvection in
the sunspot umbra by \citet{schussvog:06} and \citet{rempel:09} correctly
predict and reproduce the observed substructures -- dark lanes in CUDs
and tails attached to PUDs.

\begin{acknowledgements}
We express our thanks to C.~M\"ostl, R.~Kever, and R.~Henderson for
assisting in the observations. We are grateful to J.~A.~Bonet and
S.~Vargas Dom\'inguez for helping us to get familiar
with the MFBD algorithm, and to A.~de~Vicente for his support on
the Condor workload management system
({\tt \scriptsize http://www.cs.wisc.edu/condor/}). The authors thankfully
acknowledge the technical expertise and assistance provided by
the Spanish Supercomputing Network (Red Espa\~nola de Supercomputaci\'on),
as well as the computer resources -- the La Palma Supercomputer, located
at the Instituto de Astrof\'isica de Canarias. This work was supported by
grant IAA~300030808 of the Grant Agency of the Academy of Sciences of
the Czech Republic (AS~CR), by the Research Plan AVOZ~10030501 of AS~CR,
by the OPTICON Trans-National Access Programme, and by the
Spanish Ministerio de Educaci\'on y Ciencia through the project
ESP~2006-13030-C06-01. The SST is operated by the Royal Swedish
Academy of Sciences in the Spanish Observatorio del Roque de los
Muchachos of the Instituto de Astrof\'\i sica de Canarias.
\end{acknowledgements}

%
\bibliographystyle{aa} 

\end{document}